\documentclass[twocolumn,pre,showpacs]{revtex4-1}
\usepackage[pdfpagemode=None,pdfstartview=FitH]{hyperref}
\usepackage{hyperref}
\usepackage{graphicx}
\usepackage{epstopdf}
\usepackage[usenames]{color}

\def\br{{\bf r}}
\def\bu{{\bf u}}

\begin{document}
\title{The mechanics of a microscopic mixer: microtubules and cytoplasmic streaming in Drosophila oocytes}

\author{J. M. Deutsch}
\author{M. E. Brunner}
\affiliation{Department of Physics, University of California, Santa Cruz CA 95064}
\author{William M. Saxton}
\affiliation{Department of Biology, University of California, Santa Cruz CA 95064}

\begin{abstract}
Large scale motion of cytoplasm called cytoplasmic streaming  occurs in some large eukaryotic
cells to stir the cell’s constituents. In
{\em Drosophila} oocytes,
microtubules have been observed to undergo undulating motion, curving
to form travelling waves during cytoplasmic streaming. Here we show that this wave-like motion can
be understood physically as due to the hydrodynamic drag of streaming impellers
attached to kinesin motors moving toward the plus-ends of microtubules
whose minus ends are anchored to the cell cortex.  The tangential
forces applied to such microtubules by kinesin give rise to
bending and leads to chiral symmetry breaking causing the microtubules to propagate long travelling waves.
The waves are reminiscent of those seen in flagellar motion but of a much longer time
scale and by a different physical mechanism.  We show
how kinesin movement can produce a bulk flow of  cytoplasm surrounding a
microtubule with the range of flow greatly enhanced by the effect of
hydrodynamic coupling between impellers.  That is, a relatively small number of motors can
move a large amount of fluid. The chaotic nature of the fluid motion of
cytoplasm caused by kinesin movement along constantly changing microtubule
trajectories is important as it greatly enhances the efficiency of mixing.
Existing data on {\em in vitro} microtubule gliding assays also show this chiral instability in
two dimensions and an analysis of this gives quantitative estimates for the forces exerted by motors
and the drag coefficient.
\end{abstract}

\maketitle

\section{Introduction}

Microtubules are flexible hollow polymers of tubulin subunits that
serve many critical functions in eukaryotic cells. They are utilized
in structural contexts, because of their relatively stiff, yet flexible,
mechanical properties.  They also act as directional highways through the
viscous cytoplasm. Molecular motor proteins carry cellular constituents
along the microtubules with kinesin moving toward their fast growing
``plus-ends" and dynein moving toward their slow growing ``minus-ends".
Many studies have focused on motor driven transport processes that
generate asymmetric distributions of specific cytoplasmic constituents;
asymmetries that are essential for complex cellular functions.  In this
letter we will analyze a surprising role for microtubules and kinesin in
a mass transport process called cytoplasmic streaming that has evolved
to accomplish just the opposite; efficient homogeneous mixing of the
contents of a cell~\cite{SerbusSaxton} and see there Supplemental Movie 13~\cite{Movie13}.

Vigorous streaming is initiated during the final stages of
development of {\em Drosophila} oocytes to disperse asymmetrically
distributed mRNA particles, protein complexes, and membranous
organelles.  The mixing process is important for subsequent
embryonic development, and cannot be accomplished by
diffusion alone.  For example  a $ 1\mu m$ yolk-filled vesicle in
the cytoplasm, assuming a viscosity 8 times that of water ~\cite{LubyPhelps}
would take approximately a week to diffuse the $500 \mu m$  length
of an oocyte. This is far too long to satisfy the need for the mixing
of yolk-filled oocyte cytoplasm with the mass of yolkless nurse
cell cytoplasm that floods the anterior of the oocyte near the
end of its development.

The problem of mixing in small systems, such as in microfluidics
chambers, has been the subject of much investigation ~\cite{Squires}. The
way that a fluid at low Reynolds number is stirred has a
profound effect on the efficiency of homogenization. For
example,  the steady state flow fields generated by a single
stir bar inside a closed chamber are far less efficient than
the more chaotic flows generated when  several stir bars are used~\cite{Aref,Aref2000}.
Rigorous analyses in two dimensions show that mixing is
most efficient when topological chaos is created by three or more
stirrers, as can be shown by application of the Thurston-Nielsen
classification theorem~\cite{Thurston,Fathi,Handel}.  With this in mind, it is
interesting to note that during fast cytoplasmic streaming in
{\em Drosophila} oocytes, microtubules appear to be locally aligned
along dynamically changing curved paths that produce travelling
waves~\cite{SerbusSaxton}.  
The streaming cytoplasmic fluid moves along those paths, in patterns reminiscent
of flowing water and seaweed.
Yolk particles in the
cytoplasm have a speed of roughly $0.25 \mu m/s$.  The particles
within a region stream for the most part in one direction,
but with a non-negligible deviation in that direction over
time. The fluctuating directions, which parallel the curved paths of
the microtubules in the same region, serve to stir cytoplasm
in a chaotic manner that, as the preceding paragraph suggests,
is important for efficient mixing.

At first sight, it might appear that the time-dependent wave-like
motion is due to turbulence of the surrounding fluid. However at such minuscule Reynolds
numbers, inertial effects are negligible and turbulence is
impossible~\cite{BergRandomWalksinBiology}. Therefore one is left with a mystery of
the relationship of fluid and filament and
how such chaotic patterns could come about. The plus-end
directed motor kinesin-1 plays a crucial role in this, as
shown by 
genetic mutations in its force producing subunit (Khc)  that prevent streaming and mixing~\cite{SerbusSaxton}.
The speed of unloaded kinesin along the microtubule has been
measured to be in the $0.5 ~-~ 1 \mu m/s$ range~\cite{SvobodaBlock,MeyhoferHoward}.  This is higher
than the fluid speeds measured during fast cytoplasmic streaming~\cite{SerbusSaxton}
and is also consistent with the role of kinesin in powering
the mixing.
Inhibition of the opposing minus-end directed motor protein,
dynein, has a complementary effect,
actually stimulating fast cytoplasmic streaming ~\cite{SerbusSaxton}.  Microscopy studies
suggest that microtubules participating in this motion have their
minus ends attached to the cortex and their plus ends away from
the cortex in the interior of the oocyte~\cite{SerbusSaxton,ChaSerbus}.  

The explanation
that we analyze for the streaming phenomena is very simple: the
mass motions of cytoplasm and the microtubule undulations are complementary
physical consequences of kinesin moving cargo 
toward the plus ends of microtubules whose minus ends
are in contact with the cortex. The cargoes serve as impellers that
both drive the fluid motion away from minus-ends and generate
tangential forces that move plus-ends toward minus-ends causing
bends in the microtubules. This has been suggested previously to
explain cytoplasmic streaming, but without a physical model of the
mechanism~\cite{SerbusSaxton}. The analysis below shows that long range hydrodynamic
forces couple individual impellers, resulting in an effective
mechanism for drag-induced bulk movement of cytoplasm.  We then
show that an instability in the dynamics leads to chiral symmetry breaking giving rise to
wave-like motion of microtubules, and show that the time and length scales
predicted are in good agreement with the previous experimental
results.

\begin{figure}[htp]
\begin{center}
\includegraphics[width=\hsize]{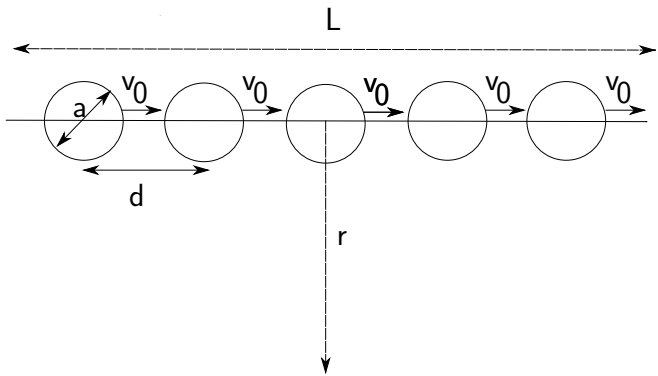}
\caption{ 
A train of spherically shaped impellers of diameter $a$ and separation $d$, all moving in a fluid with a velocity $v_0$.
The velocity is measured at a point a distance $r$ from the axis.
}
\label{fig:spheres}
\end{center}
\end{figure}

\section{Enhancement of Streaming Due to Hydrodynamics} 

Consider impellers to be objects each with maximum linear dimension
$a$, and with a mean spacing of $d$ arranged in a straight line as shown in Fig ~\ref{fig:spheres}. 
These impellers are pictured as spheres, but hydrodynamics is not sensitive to
the exact shape of an impeller, as it depends mainly on its maximum linear dimension~\cite{BergRandomWalksinBiology}. 
We will
now analyze the amount of streaming due to motion of these impellers
moving on a single microtubule. We initially consider the impellers to be much larger than
the kinesin molecules so that they dominate the hydrodynamical response
of the fluid.

If spherical impellers were close-packed along the microtubule, that is $a \approx
d$, then this problem would be equivalent to a single rod of length $L$
moving at constant velocity $v_0$ in the fluid in a direction parallel
to its long axis.  In that case, the velocity field for distances
$r \ll L$ has only a weak logarithmic dependence of $r$, meaning that up
to a correction of order $\ln(L/a)$, the velocity field is only weakly
dependent on distance and of order $v_0$. (Here we take the velocity of the fluid
to go to zero far from the rod.)
This is related to the well
known result that the drag on a rod of length $L$ is of the same order
as that of a sphere of diameter $L$ despite the latter's much greater volume~\cite{BergRandomWalksinBiology}. Such a system of densely packed impellers would
be very efficient at driving fluid motion and only a low density of microtubules
would be needed for cytoplasmic streaming.

Now consider the more realistic case in which the impeller size is less than
the spacing between them, which is much less than the length of a
microtubule, that is $a < d \ll L$.  At a distance $r \gg d$, the
velocity field will behave just as in the closed-packed case except appear
to have a diminished impeller velocity. That is for $d \ll r \ll L$,
the magnitude of the velocity $v(r)  = v_0 f(a/d) g(r)$, where $g(r)$
contains all the distance dependence of the velocity field, and $f(a/d)$
is how the velocity scales with the ratio of $a/d$. In the limit where
$a/d$ is very small, the system becomes dilute and the velocity between 
impellers will decay to zero. We recover the motion of isolated impellers
in this case, where it is well known (e.g. Stokes' drag) that the velocity
field is proportional to $a$. Therefore for small argument $x$, $f(x)$ is
linear (in other words, the fluid velocity is proportional to $a$.) 

Using this general argument we conclude that, independent of the exact shape of the impellers,
the flow velocity is reduced from the closed-packed case by a factor $\sim a/d$. The effect
of the impellers only starts decreasing substantially at a distance of order the
length of the microtubule $L$, below which it should only have a weak logarithmic
dependence. In other words, for a spherical region just enveloping a microtubule,
the fluid velocity is slowly varying and reduced from the kinesin motor velocity 
by a factor of order $a/d$.

Of course there are  many microtubules in these cells. To understand how this
affects the above analysis, consider all space filled with an infinite forest of them all oriented
in the same direction. 
First if we ignore the microtubules and just consider the spherical impellers, then
if we move to a reference frame moving with the impeller velocity, the system is static
and the velocity everywhere is zero. Therefore in the original reference frame, the
fluid is also moving uniformly at the impeller velocity. This is not correct because
we have ignored the hydrodynamic drag of the microtubules represented by the line going through
the spheres in Fig. \ref{fig:spheres}. To estimate their effect on the fluid velocity, 
denote the drag coefficient on an impeller by $b_I$ and that of a section of microtubule length $d$ by $b_M$. Then
we go to a reference frame velocity $v$ such that the total force acting on the combined system of
impellers and microtubules is zero, so that $b_I(v_0 - v) - b_M v = 0$, or $v = b_I v_0/(b_I + b_M)$.
Because the net force acting on this system is zero in this frame, $v$ is the velocity
of the fluid far from the microtubules.
Because within logarithmic corrections, the drag coefficients are proportional to the maximal
linear dimensions, then a conservative estimate of
this speed is of order $v_0 a/d$. The exact formula depends on the shape of
the impellers. This argument assumes an infinite volume of
microtubules but the corrections to this due to the finite nature of the system
are not important for the estimates we are making.

The identities of impellers in this system are still unknown but there are many possible
candidates. Anything with  large linear dimensions in at least one direction 
will give a large hydrodynamic radius $a$~\cite{BergRandomWalksinBiology}. The
other requirement is that it can attach to kinesin.  In fact
it is possible that the impellers in this situation could themselves be microtubules that are
not attached to the cortex~\cite{WangRiechmann,Seeger}.
Experimental estimates of the cytoplasmic streaming velocity
are approximately $0.25 \mu m/s$, about $\frac{1}{4}$ to $\frac{1}{2}$ the typical velocity of a kinesin
molecule. This suggests that $a/d$ is $\frac{1}{4}$ or greater.
For example, if we take the maximum dimensions of an impeller to be $250 nm$,
this predicts a spacing between impellers of $1 \mu m$ or less.

Another important biological issue is the necessity to have some microtubules in direct physical contact
with the cortex of the
oocyte, for example by tethering or by frictional forces. A free floating microtubule with kinesin moving on it will apply {\em zero} net force
to the fluid. This is a simple consequence of Newton's third law, or equivalently, conservation
of momentum. This does not contradict
the fact that bacteria are able to swim: the force propelling the bacterium forward is
countered by an equal and opposite force on the environment, leading to velocity fields that
are dipolar at large distances. Unlike the case analyzed above, this will not lead to long range hydrodynamic motion
of the fluid and will not lead to efficient cytoplasmic streaming by relatively few motor proteins. 
Contact with the cortex is crucial as it allows for transfer of force from outside of
the oocyte to the cytoplasm enabling fast streaming of the bulk to be powered by a much smaller volume of kinesin driven
impellers.

\section{Travelling Wave Instability of Microtubules}

We now show how the kinesin generated tangential forces on microtubules give rise to travelling wave conformations 
and calculate their angular and spatial frequency.
A microtubule has a configuration $\br(s)$ parameterized by 
an arclength $s$ and, at long enough length scales, can be modeled as being inextensible, 
that is $|\partial \br/\partial s| = 1$
with an elastic bending constant $C$. The inextensibility is enforced by a position dependent tension $T(s)$. 
There is also a force acting on the microtubule as a result of kinesin walking along it.
The magnitude of the force is proportional to the local speed and the size of the kinesin-driven impeller
and the direction of the force is tangent to the microtubule,
which, we will see, has the effect of making it buckle. We denote this with a force per unit length of $f_k$. 
We also include a force due to the cytoplasm streaming at a velocity $v_s$ which
we take to be in the $\hat k$ direction away from the minus end and this force tends to straighten the microtubule. This leads to the equation
\begin{equation}
\label{eq:microtubule}
\nu \frac{\partial \br}{\partial t} =  -C \frac{\partial^4 \br}{\partial s^4} + \frac{\partial}{\partial s}(T(s)\frac{\partial \br}{\partial s}) -
f_k \frac{\partial \br}{\partial s} + \nu v_{s}{\hat k} .
\end{equation}
where $\nu$ is a hydrodynamic drag coefficient per unit length. 

\begin{figure}[htp]
\begin{center}
\includegraphics[width=\hsize]{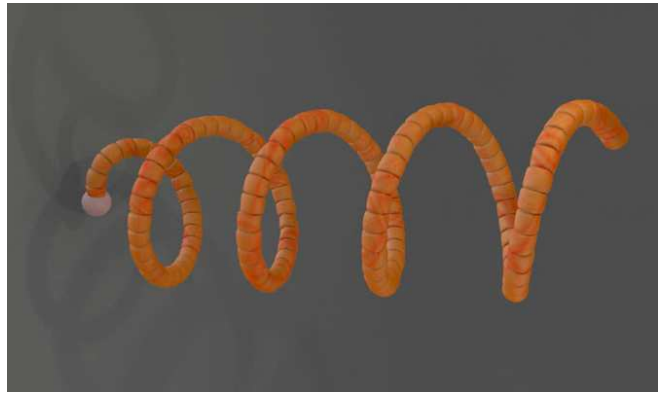}
\caption{ 
The motion of a microtubule with a constant force being applied tangent to its axis. 
A constant velocity in the horizontal direction  from left to right has also been applied representing
the background from cytoplasmic streaming. The white ball on the left represents
the point at which the microtubule's minus end is tethered. See supplemental
movie 1~\cite{SupplMovies}. In a real oocyte, the radius of curvature would
correspond to approximately $40 \mu m$ and the dimensions of the entire figure are
roughly comparable to the oocyte. The thickness of the microtubule has been magnified
by a factor of about $2000$ to facilitate viewing.
}
\label{fig:simulation}
\end{center}
\end{figure}

We first implemented this equation numerically for a range of parameters
and enforced boundary conditions that tethered the minus end against the cortex, while
the plus end was free. Starting from random initial conditions, the equation rapidly goes to a steady state that
typically is described by a curve that asymptotically becomes helical for large $s$ and rotates uniformly at constant
angular velocity. The results of a steady state configuration are shown in Fig. \ref{fig:simulation}. 
The supplemental movies~\cite{SupplMovies} discussed below, show microtubule solutions for different
parameters.
Supplemental movie 1 shows the full time dependence~\cite{SupplMovies}. The chirality of the helix depends on its initial conditions.
Therefore the direction of rotation is random but stable once steady state is reached.

As the microtubule is made longer, the angular velocity and radius of the helix go to
a constant limit. However the tethered part is not helical but nevertheless rotates
in synchrony with the rest of the microtubule. These dynamics we analyzed in detail
(see the supplemental information) to find the form of the solution to this equation.
In particular we show that this equation supports travelling waves and find the
relationship between the angular velocity of rotation $\omega$ and the asymptotic
radius of the helix, and the external velocity fields $v_s$. In the case where $v_s = 0$, the
relationship simplifies to 
\begin{equation}
\label{eq:Romega}
\nu R \omega/f_k = 1.
\end{equation}
independent of chain length.

It is interesting to note that for travelling waves, solutions can be at any scale. There is a continuous family of solutions
all with the same shape but different scale factors. In the case of a helix, many different radii $R$ are solutions to
these equations.

What determines the value of $R$ that is selected? As with other problems in pattern formation such as the ``geometric model"~\cite{Kessler}
or the full dendrite problem~\cite{Barbieri}, it is the boundary conditions
that are responsible for the unique value of $R$ that is selected. In this case, the microtubule minus end is tethered, $\bu(0) = 0$
but the plus end is free.
The solution will only exist for discrete values
of $\beta \equiv C/(R^3 f_k)$. Numerical analysis gives, $\beta = 0.05 \pm 0.0005$. 
This implies that 
\begin{equation}
\label{eq:R}
R =  (C/(\beta f_k))^{1/3}.
\end{equation}

This analysis was extended to non-zero $v_s$ and as shown in the supplemental materials~\cite{SupplMat}, does not
appreciably change the estimate we will give below for microtubule wave parameters in fast streaming oocytes.

When the microtubule is tethered to an impenetrable surface (in this case, the oocyte cortex) and the external cytoplasmic streaming velocity $v_s$
is parallel to that surface, the form of the solution changes considerably. Numerical results show that the microtubule becomes completely two dimensional,
lying close to the surface. For low enough $v_s$, travelling wave
solutions are close to prolate cycloids, meaning that the microtubule periodically loops back
on itself (see supplemental movie 2~\cite{SupplMovies}. This can be understood analytically. For sufficiently large $v_s$ it transitions to other states finally becoming 
two dimensional and looking close to a sinusoid (see supplemental movie 3~\cite{SupplMovies}. This is analyzed in detail in
the supplemental materials~\cite{SupplMat}.

We have not included hydrodynamic and steric interactions between
different microtubules except in the approximate way of giving rise to
a constant cytoplasmic streaming velocity. Given the microtubule density in the
oocyte, we expect these interactions to be substantial. However the wave
like motion that we find is quite robust. Attaching microtubules together or
considering additional forces still leads to periodic or sometimes chaotic
motion, still at the same characteristic time and spatial scales. Therefore
despite the simplicity of the model, we expect that the basic length
and time scales that we predict should be quite robust.

\section{Comparison with Experiment and Discussion}

We now check to see whether the above model is consistent with experimental
results.  

Estimates of the microtubule elastic constant $C$ vary
considerably~\cite{Felgner,GittesRigidity}  but range mostly within 
$2$ to $4 \times 10^{-23} N m^2$. We estimate the force due to the kinesin
per unit length, $f_k$ to be its velocity $v_k$ times the cytoplasmic
viscosity. We will assume as suggested by our above analysis, that
$a/d$ is between $1/4$ and $1$. We will take the kinesin velocity $v_k$ to be approximately $1
\mu m/s$~\cite{SvobodaBlock,MeyhoferHoward}. The effective viscosity of the cytoplasm for small
particles has been studied extensively and appears to vary depending on the type of cell on the length scale~\cite{LubyPhelps}. 
Particles of different sizes diffusing in the cytoplasm diffuse as if the medium had a different viscosity. Its viscoelastic properties will depend on many
factors such as the state of gelation of actin filaments~\cite{YinStossel}.
We will assume that during fast cytoplasmic streaming, the cytoplasmic actin is mainly in a "sol" state allowing fast streaming
to more readily take place. If this is not the case, the effective viscosity could be much higher~\cite{LubyPhelps}
but the dissipation would increase proportionally requiring a higher energy input. 
For small particles of radius approximately $20 nm$, the effective  viscosity 
as measured by diffusion is approximately 8 times that of water~\cite{LubyPhelps} and we will use this
value for comparison with experiment.  
This gives $f_k = (a/d) v \eta$ in the range  $2$ to $8 \times 10^{-9} N/m$.
Plugging these numbers into Eq. \ref{eq:R} gives $R$ in the range  $30$ to $50 \mu m$.
We can estimate the angular velocity using Eq. \ref{eq:Romega} (which
ignores cytoplasmic streaming). To get the largest range of times, we assume that the hydrodynamic drag
coefficient is due solely to the impellers, so that $\nu = (a/d) \eta$. This
is equivalent to $\omega = v_k/R$ or a period of $ T = 2 \pi R/v_k$ which ranges 
from $190$ to $310 s$.

Serbus and colleagues have observed microtubule behavior in living Drosophila oocytes using GFP-tubulin 
and confocal fluorescence microscopy~\cite{SerbusSaxton}. 
Time-lapse movies showed bright fibrous fluorescence, representing
groups of microtubules, in a background of diffuse fluorescence from non-polymerized GFP-tubulin.
The microtubule patterns changed over time, as did the patterns of motion of organelles that
either excluded the GFP-tubulin or were themselves auto-fluorescent. Bends in the microtubules were
at times visible in the $x-y$ optical plane and some remained visible within that plane for $30-90 s$,
allowing measurement of a radius of curvature that we expect to approximately correspond to the
$R$ in the above analysis. $R$ was measured  to be $19.5\pm 6.5 \mu m$ with $9$ measurements (standard error $2.1 \mu m$),  and
the wave velocity was $v = 0.265 \pm 0.04 \mu m/s$ with $4$ measurements. Because these waves
appear roughly sinusoidal, we can also estimate the period $T$. Assuming a wavelength of $4R$
then the measured velocity gives a period of $T= 294 s$ with an even larger error bar
considering the fact that this assumes a waveform that is probably not accurate. Nevertheless
it suggests a characteristic time.

The motion studied here can be contrasted with ciliary
motion~\cite{Kennedy} which has a period on the order of $.05 s$. Clearly
the two mechanisms have fundamentally different explanations. 
The experimental time scale points to the mechanism described here rather
than ciliary motion.  The agreement found with experiment in our above
analysis is to some extent fortuitous, given the large uncertainties
in the experimental system. But nevertheless, it provides evidence that
the simple mechanism proposed is the origin of the behavior seen in
cytoplasmic streaming.  The experimental evidence~\cite{SerbusSaxton} that dynein inhibits
streaming and that minus ends of microtubules are in contact with the
cortex~\cite{ChaSerbus}, both support this hypothesis as well. 

More generally, the above analysis elucidates, at a qualitative level, the phenomena seen
in experiment~\cite{SerbusSaxton}, of strong cytoplasmic streaming
occurring concomitantly with wave-like motion of microtubules.
The movement of a relatively few number of kinesin motors on a microtubule 
couples to the fluid causing a bulk hydrodynamic flow. The force a
kinesin exerts on a microtubule sets the microtubule into motion causing
it to execute wave-like motion. This in turn causes the flow lines around microtubules to be time dependent,
making the microtubules act as stirrers for the surrounding fluid, which leads to 
chaotic flows and strong mixing~\cite{Aref,Aref2000}.

\begin{figure}[htp]
\begin{center}
(a) \includegraphics[width=0.3\hsize]{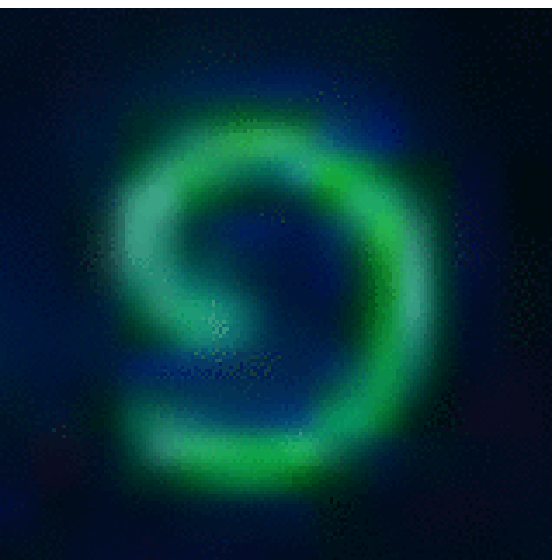}
(b)\includegraphics[width=0.3\hsize]{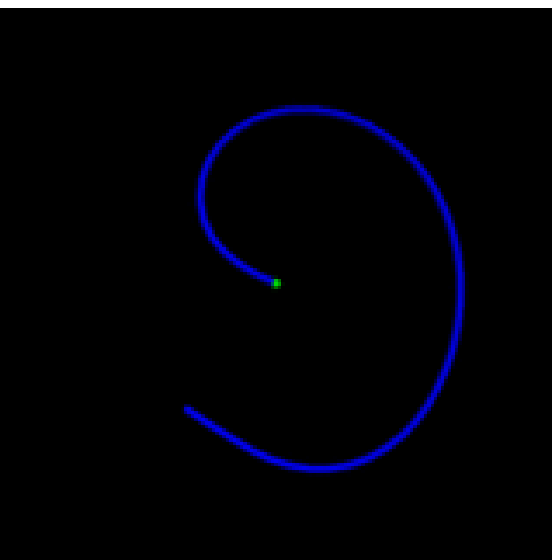}
\caption{ 
The configuration a rotating microtubule in a gliding assay~\cite{MaloneyHerskowitzKoch,MaloneyKochVideo} with the leading end
stuck to a kinesin protein is shown in (a). (b) Shows the two dimensional simulation
of Eq. \ref{eq:microtubule} showing a similar shape.
}
\label{fig:squiggle}
\end{center}
\end{figure}

Can such motion be directly observed in an {\em in vitro}
experiment?  In fact it has been seen frequently as an unwanted artifact
of sample preparation.
Microtubule gliding assays were pioneered more than two
decades ago~\cite{ValeSchnappEtAl} and have become a
standard technique to better understand the motion of kinesin motors.
When a solution microtubules and ATP is placed on a glass surface on
which there is a high density of kinesin molecules, the kinesins propel
the microtubules with their minus ends leading.  Occasionally the minus
end sticks, perhaps to damaged kinesins, and continued forces of the
trailing portion of the microtubule cause flexing and curve generation.

Microtubule gliding assay videos of high quality~\cite{MaloneyHerskowitzKoch} provide an excellent
source of data for investigating the two dimensional version of the problem
studied here~\cite{MaloneyKochVideo}. The microtubule is being pushed by kinesin molecules
that on average, apply a force tangent to the microtubule. When an
end is tethered by sticking to a kinesin molecule, as noted above, this leads to 
a physical situation that is well described by Eq. \ref{eq:microtubule}. It should be
emphasized that the values of the tangential force $f_k$ and the drag
coefficient $\nu$ will be much larger than in the oocyte case.
Consequently, sometimes microtubules flex,, 
forming a G-shaped curve that rotates around the anchored minus end 
of the microtubule.  Fig. \ref{fig:squiggle} shows a comparison of
snapshots of the microtubule with that of a
simulation of Eq. \ref{eq:microtubule}. The resemblance is quite striking and provides evidence that
the theoretical modelling of this bending instability is valid. Note that
the shape obtained is independent of the values of model parameters. From
measuring the radius $R$ of the G-like configuration and its angular
velocity of rotation, we can compute both $f_k$ and $\nu$ from Eqs.
\ref{eq:R} and \ref{eq:Romega}, giving  $\nu = \omega C/(\beta R^4)$
and $f_k = \nu R \omega$. Measurements give $R \approx 1.33 \pm 0.2
\mu m$ and the period $T = 12 \pm 1 s$. Therefore $\nu = 133 N s/m^2$
and $f_k =  9.3\times 10^{-5} N/m$, both estimates are accurate within
a factor of 2 assuming that $C$ has been well determined.  The force
exerted by a kinesin motor in this situation is close to the experimentally determined stall
value of approximately $f_m = 5 pN$~\cite{MeyhoferHoward}. Assuming that
each motor attached can exert this force, this gives
an average distance between motors of $f_m/f_k = 53 nm$. This assumes that
the attachment of the motors is flexible enough to allow microtubule interactions capable of exerting this stall force. If
this is not the case, we expect the kinesins to be spaced more closely but
by no more than half this amount. This distance
seems reasonable given the high density of kinesin molecules attached
to the glass, and the size of the $\alpha-\beta$ tubulin dimer which is
approximately $8 nm$ in length. Thus the above considerations can give
a quantitative measure of the forces and drag in these gliding assays.

Nevertheless, it would also be of interest to verify the travelling wave solutions experimentally
in three dimensions for isolated
microtubules and kinesin, for example, by utilizing a single molecule optical trap, or by some other
means, allowing for a more rigorous test of the predictions made here.

It is of interest to speculate on the reason for this microscopic mixing
mechanism.  If the microtubules were attached from the {\em plus} end,
there would be no dynamical instability from kinesin generated forces
but the cytoplasmic streaming would be more efficient, with greater transfer
of force from outside the oocyte, because the tension at the
tether point would increase linearly with the length of the microtubule,
whereas the tension saturates to a constant value in the case considered
above. However as discussed earlier, complex stirring patterns are
expected to be far more efficacious in mixing than steady-state flows.
This agitation could then serve an important biological function.

In summary, we have found a very simple set of conditions for generating efficient mixing
in large eukaryotic cells. If microtubules are aligned so that their minus ends
are in contact with the cell perimeter and dynein is suppressed, this will lead 
(a) to cytoplasmic streaming in the bulk of the cell and (b) wave-like motion of
microtubules. The latter will promote efficient mixing of the cell's contents by
inducing chaotic flows.
The efficiency of the cytoplasmic streaming is due to the long range nature of
hydrodynamic coupling and the indirect linkage of microtubules to the oocyte surroundings via the cortex, leading to momentum transfer.
The wave-like motion is due to an instability in the dynamics which breaks chiral
symmetry and is a consequence of the tangential forces exerted by motor proteins on the elastic
microtubules. Given the uncertainties in the experimental system and approximations
made in the analysis, The wave-like motion seen in experiments agrees well with our theoretical predictions
suggesting these effects represent a robust phenomenon.
We are now engaged in experimentation to gather more accurate data on microtubule behavior 
{\em in vivo} and {\em in vitro} under various conditions in order to further understand
the mechanisms for fast cytoplasmic streaming discussed here.

The authors would like to thank Ian Carbone, and Bill Sullivan for useful discussions.
We also would like to gratefully acknowledge Andy Maloney and Steven J. Koch for their permission
to use a snapshot of their video, Fig.  \ref{fig:squiggle}(a) \cite{MaloneyKochVideo} of their
experimental results~\cite{MaloneyHerskowitzKoch}. This was generously distributed to us
by them~\cite{KochLab} through the practice of Open Notebook Science~\cite{OpenNotebookScience} 
This material is based upon work supported by National Institutes of Health GM046295 (to W.M.S.), National Science Foundation
CCLI Grant DUE-0942207 (to J.M.D.),
the  Defense Threat Reduction Agency basic research grant HDTRA-1-09-1-0018 (to Steven J. Koch),
and the National Science Foundation IGERT Grant DGE-0549500 (to Steven J. Koch).

{\em Note:} After the completion of this work, we came across work of Bourdieu
et al.~\cite{Bourdieu95} which analyzed their experimental results on
spirals in  mysosin and kinesin gliding assays using the two dimensional
version of Eq. \ref{eq:microtubule}, obtaining the identical scaling,
and simulating it using a nearly identical approach. This provides
further evidence that the mechanism we have proposed for cytoplasmic
streaming is physically viable.

\end{document}